%% file: tark07.tex
\documentclass[]{article}
\usepackage{proceed2e}

\def\smallromani{\renewcommand{\theenumi}{\roman{enumi}}
\renewcommand{\labelenumi}{(\theenumi)}}

\newcommand{\Proof}{\NI
                    {\bf Proof.}\ }

\usepackage{url}
\usepackage{amsmath}
\usepackage{handbookbib}

\usepackage{color}
\usepackage{graphics}

\include{ecmds}
\usepackage{amsfonts}
\usepackage{latexsym}
\usepackage{alltt}

\newtheorem{theorem}{Theorem}%[section]
\newtheorem{defined}{Definition}
\newenvironment{definition}{\begin{defined} \rm}{\end{defined}}
\newtheorem{exa}{Example}

\newtheorem{lemma}{Lemma}

\newtheorem{note}{Note}

\title{Epistemic Analysis of Strategic Games with Arbitrary Strategy Sets}

\author{{\bf Krzysztof R. Apt} \\
CWI, Amsterdam, the Netherlands \\
and University of Amsterdam\\
}

\begin{document}

\date{}
\maketitle
%\doublespace

\begin{abstract}
  We provide here an epistemic analysis of arbitrary strategic games
  based on the possibility correspondences.  Such an analysis calls for the use
  of transfinite iterations of the corresponding operators.  Our
  approach is based on Tarski's Fixpoint Theorem and applies both to
  the notions of rationalizability and the iterated elimination of
  strictly dominated strategies.
\end{abstract}

\section{Introduction}

Epistemic analysis of strategic games (in short, games) aims
at predicting the choices of rational players in the presence of
(partial or common) knowledge or belief of the behaviour of other players.  Most
often it focusses on the iterated elimination of never best responses
(a notion termed as rationalizability), and the iterated elimination
of strictly dominated strategies (IESDS).

Starting with \cite{Aum87a}, \cite{BD87} and \cite{TW88} a large body
of literature arose that investigates the epistemic foundations of
rationalizability by modelling the reasoning employed by players in
choosing their strategies.  Such an analysis, based either on
possibility correspondences and partition spaces,
or Harsanyi type spaces, is limited either
to finite or compact games with continuous payoffs, or to two-player
games, see, e.g., \cite{BB99} or \cite{EP06}.

In turn, in the case of IESDS the epistemic analysis has focussed on
finite games (with an infinite hierarchy of beliefs) and strict
dominance either by pure or by mixed strategies, see, e.g.
\cite{BFK04}.

In this paper we provide an epistemic analysis of arbitrary strategic
games based on the possibility correspondences.
More specifically, denote by $\textbf{RAT}({\overline{\phi}})$
 the property
that each player $i$ uses a monotonic property $\phi_i$ to select
his strategy (`each player $i$ is $\phi_i$-rational'). Then the
following sets of strategy profiles coincide:

\begin{itemize}

\item those that the players choose in the states in which $\textbf{RAT}({\overline{\phi}})$
is common knowledge,

\item those that the players choose in the states in which $\textbf{RAT}({\overline{\phi}})$
is true and is common belief,

\item those that remain after the iterated
elimination of the strategies that are not $\phi_i$-optimal.  

\end{itemize}

This requires that transfinite iterations of the strategy elimination
are allowed and covers the usual
notion of rationalizability and a 'global' version\footnote{The
  concepts of `global' and `local' versions are clarified in Section
  \ref{sec:setup}.} of the iterated elimination of strictly dominated
strategies.  For the customary, `local' version of the iterated
elimination of strictly dominated strategies (that is defined using a
non-monotonic property) we justify the statement
\begin{quote}
  common knowledge of rationality implies that the players will choose
  only strategies that survive the iterated elimination of strictly
  dominated strategies
\end{quote}
for arbitrary games and transfinite iterations of the
elimination process. Rationality refers here to the concept studied in
\cite{Ber84}.

Our results complement the findings of \cite{Lip91} in which
transfinite ordinals are used in a study of limited rationality and
\cite{Lip94}, where a two-player game is constructed for which the
$\omega_0$ (the first infinite ordinal) and $\omega_{0} + 1$ iterations of
the rationalizability operator of \cite{Ber84} differ.  In turn,
\cite{HS98} show that in general arbitrary ordinals are necessary in
the epistemic analysis of strategic games based on the partition
spaces.  Further, as argued in \cite{CLL05}, the notion of IESDS \`{a}
la \cite{MR90}, when used for arbitrary games, also requires
transfinite iterations of the underlying operator.

The relevance of monotonicity in the context of epistemic analysis of 
finite strategic games has already been pointed out in \cite{vB07}, where the 
notions of strict dominance and rationalizability are studied using a public
announcement logic.

\section{Preliminaries}
\label{sec:prelim}

In this section we recall basic results concerning 
operators on a complete lattice and the relevant notions concerning strategic games.

\subsection{Operators}

Consider a fixed complete lattice $(D, \sse)$ with the largest element $\top$.
In what follows we use ordinals and denote them by $\alpha, \beta, \gamma$.
Given a, possibly transfinite, sequence $(G_{\alpha})_{\alpha < \gamma}$ of
elements of $D$ we denote their join and meet respectively by
$\bigcup_{\alpha < \gamma} G_{\alpha}$
and $\bigcap_{\alpha < \gamma} G_{\alpha}$.

\begin{definition}
Let $T$ be an operator on $(D, \sse)$, i.e., $T: D \myra D$.

\begin{itemize}

\item We call $T$ \oldbfe{monotonic} if for all $G_1, G_2$
\[
\mbox{$G_1 \sse G_2$ implies $T(G_1) \sse T(G_2)$.}
\]

\item We call $T$ \oldbfe{contracting} if for all $G$
\[
T(G) \sse G.
\]

\item We say that an element $G$ is a \oldbfe{fixpoint} of $T$ if $G = T(G)$
and a \oldbfe{post-fixpoint} of $T$ if $G \sse T(G)$.

\item We define by 
transfinite induction a sequence of elements $T^{\alpha}$ of $D$, where $\alpha$ is an ordinal, as follows:

\begin{itemize}

  \item $T^{0} := \top$,

  \item $T^{\alpha+1} := T(T^{\alpha})$,

  \item for all limit ordinals $\beta$, $T^{\beta} := \bigcap_{\alpha < \beta} T^{\alpha}$.
  \end{itemize}

\item We call the least $\alpha$ such that $T^{\alpha+1} = T^{\alpha}$ the \oldbfe{closure ordinal} of $T$
and denote it by $\alpha_T$.  We call then $T^{\alpha_T}$ the \oldbfe{outcome of} (iterating) $T$ and write it alternatively as $T^{\infty}$.
\HB
\end{itemize}
\end{definition}

So an outcome is a fixpoint reached by a transfinite iteration that
starts with the largest element.  In general, the outcome of an
operator does not need to exist but we have the following classic
result due to \cite{Tar55}.\footnote{We use here its `dual' version in
  which the iterations start at the largest and not at the least
  element of a complete lattice.}
\II

\NI
\textbf{Tarski's Fixpoint Theorem} 
Every monotonic operator $T$ on $(D, \sse)$
has an outcome, i.e., $T^{\infty}$ is well-defined.
Moreover,
\[
T^{\infty} = \nu T = \cup \{G \mid G \sse T(G)\},
\]
where $\nu T$ is the largest fixpoint of $T$.
\vspace{2mm}

In contrast, a contracting operator does not need to have a largest fixpoint.
But we have the following obvious observation.

\begin{note} \label{note:con}
Every contracting operator $T$ on $(D, \sse)$ has an outcome, i.e., 
$T^{\infty}$ is well-defined.
\end{note}

In Section \ref{sec:consequences} we shall need the following lemma.

\begin{lemma} \label{lem:inc}
Consider two operators $T_1$ and $T_2$ on $(D, \sse)$ such that
\begin{itemize}
\item for all $G$, $T_1(G) \sse T_2(G)$,

\item $T_1$ is monotonic,

\item $T_2$ is contracting.
\end{itemize}
Then $T_1^{\infty} \sse T_2^{\infty}$.
\end{lemma}
\Proof
We first prove by transfinite induction that for all $\alpha$
\begin{equation}
T_1^{\alpha} \sse T_2^{\alpha}.
  \label{equ:inc}
\end{equation}

By the definition of the iterations we only need to consider the induction
step for a successor ordinal.  So suppose the claim holds for some
$\alpha$. Then by the first two assumptions and the induction
hypothesis we have the following string of inclusions and equalities:
\[
T_1^{\alpha + 1} =   T_1(T_1^{\alpha}) \sse T_1(T_2^{\alpha}) \sse T_2(T_2^{\alpha}) = T_2^{\alpha + 1}.
\]

This shows that for all $\alpha$ (\ref{equ:inc}) holds.
By Tarski's Fixpoint Theorem and Note \ref{note:con} the outcomes of
$T_1$ and $T_2$ exist, which implies the claim.
\HB

\subsection{Strategic games}

Given $n$ players ($n > 1$) by a \oldbfe{strategic game} (in short, a
\oldbfe{game}) we mean a sequence
$
(S_1, \LL, S_n, p_1, \LL, p_n),
$
where for each $i \in [1..n]$

\begin{itemize}
\item $S_i$ is the non-empty set of \oldbfe{strategies} (sometimes called \oldbfe{actions})
available to player $i$,

\item $p_i$ is the \oldbfe{payoff function} for the  player $i$, so
$
p_i : S_1 \times \LL \times S_n \myra \cal{R},
$
where $\cal{R}$ is the set of real numbers.
\end{itemize}

We denote the strategies of player $i$ by $s_i$, possibly with some
superscripts.  Given $s \in S_1 \times \LL \times S_n$ we denote the
$i$th element of $s$ by $s_i$, write sometimes $s$ as $(s_i, s_{-i})$,
and use the following standard notation:

\begin{itemize}
\item $s_{-i} := (s_1, \LL, s_{i-1}, s_{i+1}, \LL, s_n)$,

\item $S_{-i} := S_1 \times \LL \times S_{i-1} \times S_{i+1} \times \LL \times S_n$.

\end{itemize}

Given a finite non-empty set $A$ we denote by
$\Delta A$ the set of probability distributions over $A$ and call
any element of $\Delta S_i$ a \oldbfe{mixed strategy} of player $i$.

In what follows we assume an initial strategic game
\[
H := (T_1, \LL, T_n, p_1, \LL, p_n).
\]
A \oldbfe{restriction} of $H$ is a sequence $(S_1, \LL, S_n)$ such that
$S_i \sse T_i$  for $i \in [1..n]$. We identify the restriction 
$(T_1, \LL, T_n)$ with $H$.
We shall focus on the complete lattice
that consists of the set of all restrictions of the game $H$
ordered by the componentwise set inclusion:
\[
\mbox{$(S_1, \LL, S_n) \sse (S'_1, \LL, S'_n)$ iff $S_i \sse S'_i$ for all $i \in [1..n]$.}
\]
So $H$ is the largest element in this lattice and
$\bigcup_{\alpha < \gamma}$
and $\bigcap_{\alpha < \gamma}$ are the customary set-theoretic 
operations on the restrictions.

Consider now a restriction $G := (S_1, \LL, S_n)$ of $H$
and two strategies $s_i, s'_i$ from $T_i$ (so \emph{not necessarily} from $S_i$). 
We say that 
$s_i$ \oldbfe{is strictly dominated} \oldbfe{on} $G$ by $s'_i$
if
\[
\fa s_{-i} \in S_{-i} \: p_{i}(s'_i, s_{-i}) > p_{i}(s_i, s_{-i}),
\]
and write then $s'_i \succ_{G} s_i$.

In the case of finite games, once the payoff function is extended in 
the expected way to mixed strategies, the relation $\succ_{G}$ between
a mixed strategy and a pure strategy is defined in the same way.

Further, given a restriction $G' := (S'_1, \LL, S'_n)$ of $H$,
we say that 
the strategy $s_i$ from $T_i$ is a \oldbfe{best response in
$G'$ to some belief $\mu_i$ held in $G$}
if 
\[
\fa s'_i \in S'_i \:
p_i(s_i, \mu_i) \geq p_i(s'_i, \mu_i).
\]
A belief held in $G := (S_1, \LL, S_n)$ can be 

\begin{itemize}
\item a joint strategy
of the opponents of player $i$ in $G$ (i.e., $s_{-i} \in S_{-i}$), 

\item or, in the case the game is finite, a joint mixed strategy of
  the opponents of player $i$ (i.e., $(m_1, \LL, m_{i-1}, m_{i+1},
  \LL, m_n)$, where $m_j \in \Delta S_j$ for all $j$),

\item or a \oldbfe{correlated strategy} of the opponents of player $i$
  (i.e., $m \in \Delta S_{-i}$).
\end{itemize}

Every joint mixed strategy of the opponents of player $i$
can be identified with their correlated strategy.

\section{Set up}
\label{sec:setup}

The assumption that each player is rational is one of the basic
stipulations within the framework of strategic games. However,
rationality can be differently interpreted by different
players.\footnote{This matter is obfuscated by the fact that the
  etymologically related noun `rationalizability' stands by now for
  the concept introduced in \cite{Ber84} and \cite{Pea84} that refers
  to the outcome of iterated elimination of never best responses.} This
may for example mean that a player

\begin{itemize}
\item does not choose a strategy strictly dominated by another pure/mixed strategy,
  
\item chooses only best replies to the (beliefs about the) strategies
  of the opponents.
\end{itemize}

In this paper we are interested in analyzing situations in which each
player pursues his own notion of rationality, more specifically the
situations in which this information is common knowledge or common
belief.  As a special case we cover then the usually analyzed
situation in which all players use the same notion of rationality.

Given player $i$ in a strategic game $H := (T_1, \LL, T_n, p_1,
\LL,p_n)$ we formalize his notion of rationality as a property
$\phi_i(s_i, G, G')$ that holds between a state $s_i \in T_i$ and
restrictions $G$ and $G'$ of $H$.  Intuitively, $\phi(s_i, G, G')$
holds if $s_i$ is an `optimal' strategy for player $i$ within the
restriction $G$ in the context of $G'$, assuming that he uses the property
$\phi$ to select optimal strategies.

Here are some examples of the property $\phi$ which show that 
the abovementioned rationality notions can be formalized in a number
of natural ways:

\begin{itemize}
  
\item $\emph{sd}(s_i, G, G')$ that holds iff the strategy $s_i$ of
  player $i$ is not strictly dominated on $G$ by any strategy from the
  restriction $G' := (S'_1, \LL, S'_n)$ of $H$ (i.e., $\neg \te s'_i
  \in S'_i \: s'_i \succ_{G} s_i$),

\item (assuming $H$ is finite) $\emph{msd}(s_i, G, G')$ that holds iff
  the strategy $s_i$ of player $i$ is not strictly dominated on $G$ by
  any of its mixed strategy from the restriction $G' := (S'_1, \LL,
  S'_n)$ of $H$, (i.e., $\neg \te m'_i \in \Delta S'_i \: m'_i
  \succ_{G} s_i$),
  
\item $\emph{br}(s_i, G, G')$ that holds iff the strategy $s_i$ of
  player $i$ is a best response in the restriction $G' := (S'_1, \LL,
  S'_n)$ of $H$ to some belief $\mu_i$ held in $G$ (i.e., for some
  belief $\mu_i$ held in $G$, $\fa s'_i \in S'_i \: p_i(s_i, \mu_i)
  \geq p_i(s'_i, \mu_i)$).
\end{itemize}

Two natural possibilities for $G'$ are $G' = H$ or $G' = G$.  We then
abbreviate $\phi(s_i, G, H)$ to $\phi^{\: g}(s_i, G)$ and $\phi(s_i,
G, G)$ to $\phi^{\: l}(s_i, G)$ and henceforth focus on the binary
properties $\phi(\cdot, \cdot)$.  (The superscript `$g$' stands for
'global' and `$l$' for 'local'.)

We say that the property $\phi(\cdot, \cdot)$ (used by player $i$)
is \oldbfe{monotonic} if
for all restrictions $G$ and $G'$ of $H$ and $s_i \in T_i$
\[
\mbox{$G \sse G'$ and $\phi(s_i, G)$ implies $\phi(s_i, G')$.}
\]

Each sequence of properties $\overline{\phi} := (\phi_1, \LL, \phi_n)$
determines an operator $T_{\overline{\phi}}$ on the restrictions of
$H$ defined by
\[
T_{\overline{\phi}}(G) := (S'_1, \LL, S'_n),
\]
where $G := (S_1, \LL, S_n)$ and for all $i \in [1..n]$
\[
S'_i := \{ s_i \in S_i \mid \phi_i(s_i, G)\}.
\]

Since $T_{\overline{\phi}}$ is contracting, by Note \ref{note:con} it
has an outcome, i.e., $T_{\overline{\phi}}^{\infty}$ is well-defined.
Moreover, if each $\phi_i$ is monotonic, then $T_{\overline{\phi}}$ is
monotonic and by Tarski's Fixpoint Theorem its largest fixpoint $\nu
T_{\overline{\phi}}$ exists and equals $T_{\overline{\phi}}^{\infty}$.

Intuitively, $T_{\overline{\phi}}(G)$ is the result of removing from
$G$ all strategies that are not $\phi_i$-optimal. So the outcome of
$T_{\overline{\phi}}$ is the result of the iterated elimination of
strategies that for player $i$ are not $\phi_i$-optimal, where $i \in
[1..n]$.

When each property $\phi_i$ equals $\phi$, we write $T_{\phi}$ instead
of $T_{\overline{\phi}}$.  The natural examples of such an iterated
elimination of strategies that were discussed in the literature
are:\footnote{The reader puzzled by the existence of multiple
  definitions for the apparently uniquely defined concepts is
  encouraged to consult \cite{Apt07}.}

\begin{itemize}
\item iterated elimination of strategies that are strictly dominated by another strategy;
  
  This corresponds to the iterations of the $T_{\textit{sd}^{\: l}}$
  operator in the case of \cite{DS02}) and of the $T_{\textit{sd}^{\:
      g}}$ operator in the case of \cite{CLL05}.

\item (for finite games)
iterated elimination of strategies that are strictly dominated by a mixed strategy;

This is the customary situation studied starting with \cite{LR57} that
corresponds to the iterations of the $T_{\textit{msd}^{\: l}}$
operator.

\item iterated elimination of strategies that are never best responses to some belief;

This corresponds to the iterations of the $T_{\textit{br}^{\: g}}$ operator
in the case of \cite{Ber84} and the $T_{\textit{br}^{\: l}}$ operator
in the case of \cite{Pea84}, in each case for an appropriate set of beliefs.

\end{itemize}

Usually only the first $\omega_o$ iterations of the corresponding
operator $T$ are considered, i.e., one studies
$T^{\omega_0}$, that is $\bigcap_{i< \omega_0}
T^{i}$, and not $T^{\infty}$.

In the next section we assume that each player $i$ employs some
property $\phi_i$ to select his strategies and analyze the situation
in which this information is common knowledge.  To determine which
strategies are then selected by the players we shall use the
$T_{\overline{\phi}}$ operator.  We shall also explain why in general
transfinite iterations are necessary.

\section{Two theorems}
\label{sec:epistemic}

To proceed further we need to recall some basic facts 
concerning the epistemic analysis of strategic games.
The approach taken below is based on the
partition spaces, or more generally possibility correspondences.
We follow here the exposition of \cite{BB99}.

Given the initial game $H = (T_1, \LL, T_n, p_1, \LL,p_n)$ we assume a
space $\Omega$ of \oldbfe{states} such that in each state $\omega \in
\Omega$ each player $i$ chooses the strategy $s_i(\omega) \in T_i$.
We assume that for $i \in [1..n]$ we have
$|\Omega| \geq |T_i|$, where
for a set $A$ we denote its cardinality by $|A|$.
A natural example of $\Omega$ satisfying this assumption
is the set of joint strategies in the game $H$.
Then given a state $\omega := s$ we simply have $s_i(\omega) = s_i$.

A \oldbfe{possibility correspondence} is a mapping from $\Omega$ to ${\cal P}(\Omega)$.
We consider three properties of a possibility correspondence $P$:

\begin{enumerate}\smallromani
\item for all $\omega$, $P(\omega) \neq \ES$,

\item for all $\omega$ and $\omega'$, $\omega' \in P(\omega)$ implies $P(\omega') = P(\omega)$,

\item for all $\omega$, $\omega \in P(\omega)$.
\end{enumerate}

If the possibility correspondence satisfies properties (i) and (ii),
we call it a \oldbfe{belief correspondence} and if it satisfies
properties (i)--(iii), we call it a \oldbfe{knowledge
  correspondence}.\footnote{In the modal logic terminology a belief
  correspondence is a frame for the modal logic KD45 and a
  knowledge correspondence is a frame for the modal logic S5, see, e.g. \cite{BRV01}.}

In the latter case
the correspondence $P$ yields a partition
$\{P(\omega) \mid \omega \in \Omega\}$ of $\Omega$.

We assume that each player $i$ has a
possibility correspondence
$P_i$ on $\Omega$.
Recall that an \oldbfe{event} is a subset of $\Omega$, 
and that an event $F$ is \oldbfe{evident} if for all 
$\omega \in F$ we have
$P_i(\omega) \sse F$ for all $i \in [1..n]$.

Following \cite{Aum76} if each $P_i$ is a knowledge correspondence, we
say that an event $E$ is a \oldbfe{common knowledge in the state}
$\omega \in \Omega$ if for some evident event $F$ we have $\omega \in
F \sse E$.  We write then $\omega \in K^* E$.

Finally, (using a characterization of \cite{MS89}) if each $P_i$ is a
belief correspondence, we say that an event $E$ is a \oldbfe{common belief
  in the state} $\omega \in \Omega$ if for some evident event $F$ we
have $\omega \in F \sse B E$, where $B E = \{\omega \in \Omega \mid
\fa i \in [1..n] \: P_i(\omega) \sse E\}$.  We write then $\omega \in
B^* E$.

Each event $E$ determines a restriction $G_E$ of $H$ defined by $G_E := (S_1, \LL, S_n)$,
where  for all $j \in [1..n]$
\[
S_j := \{s_j(\omega') \mid \omega' \in E\}.
\]

In particular, when player $i$ knows (respectively, believes)
that the state is in
$P_i(\omega)$, the restriction $G_{P_i(\omega)}$ represents his
knowledge (respectively, his belief) about the players' strategies.

Given now a property $\phi_i(\cdot, G)$ that player $i$ uses to select
his strategies in the restriction $G$ of $H$, we say that player $i$
is $\phi_i$-\oldbfe{rational in the state} $\omega$ if
$\phi_i(s_{i}(\omega), G_{P_i(\omega)})$ holds.  Intuitively, if the
state of the world is $\omega$, player $i$ only knows (respectively,
believes) that the state of the world is in $P_i(\omega)$. So $G_{P_i(\omega)}$
is the game he knows (respectively, believes in). Hence $\phi_i(s_{i}(\omega),
G_{P_i(\omega)})$ captures the idea that if player $i$ uses
$\phi_i(\cdot, \cdot)$ to select his optimal strategy in the game he
is `aware of', then in the state $\omega$ he indeed acts 'rationally'.

We are interested in the strategies selected by each player in the
states in which it is common knowledge (or true and common belief)
that each player $i$ is $\phi_i$-rational.  To this end we introduce
the following set of
states\footnote{$\textbf{RAT}({\overline{\phi}})$ is always used in
  the context of specific possibility correspondences.}:
\[
\mbox{$\textbf{RAT}({\overline{\phi}}) := \{\omega \in \Omega \mid $ each player $i$ is $\phi_i$-rational in $\omega$\}}
\]
and focus on the following two sets of states:
\begin{tabbing}
$CK({\overline{\phi}}) := \{\omega \in \Omega \mid $ \= for some knowledge \\
 \> correspondences $P_1, \LL, P_n$\\
 \> $\omega \in K^* \textbf{RAT}({\overline{\phi}})$\},
\end{tabbing}
\begin{tabbing}
$CB({\overline{\phi}}) := \{\omega \in \Omega \mid $ \= for some belief \\ 
 \> correspondences $P_1, \LL, P_n$\\
 \> $\omega \in \textbf{RAT}({\overline{\phi}})$ and $\omega \in B^* \textbf{RAT}({\overline{\phi}})$\}
\end{tabbing}
and the corresponding restrictions $G_{CK({\overline{\phi}})}$
and  $G_{CB({\overline{\phi}})}$ of $H$.

The following result then characterizes for arbitrary strategic games
the restrictions $G_{CK({\overline{\phi}})}$ and $G_{CB({\overline{\phi}})}$ 
in terms of the operator $T_{\overline{\phi}}$.

\begin{theorem} \label{thm:epist1}
Suppose that each property $\phi_i$ is monotonic. Then 
\[
G_{CK(\overline{\phi})} = G_{CB(\overline{\phi})} = T_{\overline{\phi}}^{\infty}.
\]
\end{theorem}
\Proof
We prove three inclusions.

\NI
(i) $G_{CK(\overline{\phi})} \sse G_{CB(\overline{\phi})}$.
\II

This inclusion (for an arbitrary $\overline{\phi}$)
is an immediate consequence of the following alternative
characterization of common knowledge due to \cite{MS89}: if each $P_i$
is a knowledge correspondence, an event $E$ is a common knowledge in
the state $\omega \in \Omega$ if for some evident event $F$ we have
$\omega \in F \sse K E$, where $K E = \{\omega \in \Omega \mid \fa i
\in [1..n] \: P_i(\omega) \sse E\}$.
\II

\NI
(ii) $G_{CB(\overline{\phi})} \sse T_{\overline{\phi}}^{\infty}$.
\II

Take a strategy $s_i$ that is an element of the $i$th component of 
$G_{CB(\overline{\phi})}$. So $s_i = s_i(\omega)$ for some
$\omega \in {CB(\overline{\phi})}$.
Then $\omega \in \textbf{RAT}({\overline{\phi}})$ and $\omega \in B^* \textbf{RAT}({\overline{\phi}})$.
The latter implies that for some evident event $F$
\begin{equation}
  \label{equ:F}
\omega \in F \sse \{\omega' \in \Omega \mid \fa i \in [1..n] \: P_i(\omega') \sse \textbf{RAT}({\overline{\phi}})\}.  
\end{equation}

Take now an arbitrary $\omega' \in F \cap \textbf{RAT}({\overline{\phi}})$ and $i \in [1..n]$.
Since $\omega' \in \textbf{RAT}({\overline{\phi}})$, 
player $i$ is $\phi_i$-rational in $\omega'$,
i.e., $\phi_i(s_i(\omega'), G_{P_i(\omega')})$ holds.
But $F$ is evident, so $P_i(\omega') \sse F$. Moreover by (\ref{equ:F})
$P_i(\omega') \sse \textbf{RAT}({\overline{\phi}})$, so 
$P_i(\omega') \sse F \cap \textbf{RAT}({\overline{\phi}})$.
Hence 
$G_{P_i(\omega')} \sse G_{F \cap \textbf{RAT}({\overline{\phi}})}$ and
by the monotonicity of $\phi_i$ we
conclude that $\phi_i(s_i(\omega'), G_{F \cap \textbf{RAT}({\overline{\phi}})})$
holds.

By the definition of $T_{\overline{\phi}}$
this means that $G_{F \cap \textbf{RAT}({\overline{\phi}})} \sse T_{\overline{\phi}}(G_{\textbf{RAT}({\overline{\phi}})})$, i.e. that
$G_{F \cap \textbf{RAT}({\overline{\phi}})}$ is a post-fixpoint of $T_{\overline{\phi}}$.
Hence by Tarski's Fixpoint Theorem 
$G_{F \cap  \textbf{RAT}({\overline{\phi}})} \sse T_{\overline{\phi}}^{\infty}$.

But $s_i = s_i(\omega)$ and $\omega \in F \cap {\textbf{RAT}({\overline{\phi}})}$, so 
we conclude by the above inclusion 
that $s_i$ is an element of the $i$th component of $T_{\overline{\phi}}^{\infty}$.
This proves $G_{CB(\overline{\phi})} \sse T_{\overline{\phi}}^{\infty}$.
\II

\NI
(iii) $T_{\overline{\phi}}^{\infty} \sse G_{CK(\overline{\phi})}$.

Recall that 
$H = (T_1, \LL, T_n, p_1, \LL, p_n)$. We first define 
\begin{itemize}
\item the functions $s_1:\Omega \myra T_1, \LL, s_n:\Omega \myra T_n$,

\item an event $E$,

\item the knowledge correspondences $P_1, \LL, P_n$.

\end{itemize}

Suppose $T^{\infty}_{\overline{\phi}} = (S_1, \LL, S_n)$. Choose $j \in
[1..n]$ such that the set $S_{j_0}$ has the largest cardinality
among the sets $S_1, \LL, S_n$. Define the
function $s_{j_0}:\Omega \myra T_{j_0}$ arbitrarily, but so that it
is onto (note that this is possible since by assumption
$|\Omega| \geq |T_{j_0}|$) and let $E :=
s^{-1}_{j_0}(S_{j_0})$.

Our aim is to ensure that 
\[
G_E = T_{\overline{\phi}}^{\infty}.
\]
So we define each function $s_k: \Omega \myra T_k$, where $k \neq j_0$, in such a way
that $s^{-1}_{k}(S_{k}) = E$. Note that this is possible since 
$|E| \geq |S_{j_0}| \geq |S_k|$.

Next, we define each knowledge correspondence $P_i$
arbitrarily but so that for all $\omega \in E$ we have $P_i(\omega) =
E$.  Then for all $i \in [1..n]$
\[
G_{P_i(\omega)} = G_E.
\]

We now show that for all $\omega \in E$ each player $i$ is
$\phi_i$-rational in $\omega$.  So take an arbitrary $\omega \in E$
and $i \in [1..n]$.  By the definition of the function $s_i(\cdot)$ a
strategy $s_i \in S_i$ exists such that $s_i = s_i(\omega)$.  Now,
$T_{\overline{\phi}}^{\infty}$ is a fixpoint of $T_{\overline{\phi}}$,
so $\phi_i(s_i, T_{\overline{\phi}}^{\infty})$ holds.  But
$T_{\overline{\phi}}^{\infty} = G_E = G_{P_i(\omega)}$, so
$\phi_i(s_i(\omega), G_{P_i(\omega)})$ holds, i.e. player $i$ is
indeed $\phi_i$-rational in $\omega$.

To complete the proof take now an arbitrary strategy $s_i \in S_i$. By
the definition of the function $s_i(\cdot)$ a state $\omega \in E$
exists such that $s_i = s_i(\omega)$.  Further, we just showed that 
each player $j$ is
$\phi_j$-rational in $\omega$.  But by the definition of the
knowledge correspondences
$E$ is an evident event, so it is common
knowledge in $\omega$ that each player $j$ is $\phi_j$-rational in
$\omega$. Hence $\omega \in CK(\overline{\phi})$ and consequently
$s_i$ is an element of the $i$th component of
$G_{CK(\overline{\phi})}$.

This proves that 
$T_{\overline{\phi}}^{\infty} \sse G_{CK(\overline{\phi})}$.
\HB
\VV

This theorem shows that when each property $\phi_i$ is monotonic, the
strategy profiles that the players choose in the states in which it is
common knowledge that each player $i$ is $\phi_i$-rational (or in
which each player $i$ is $\phi_i$-rational and it is common belief
that each player $i$ is $\phi_i$-rational), are exactly those that
remain after the iterated elimination of the strategies that are not
$\phi_i$-optimal.
It generalizes corresponding results established for finite
strategic games (for their survey see \cite{BB99}) to the case of
arbitrary strategic games and arbitrary monotonic properties $\phi_i$.

In \cite{CLL05}, \cite{Lip94} and \cite{Apt07} examples are provided
showing that for the properties of strict dominance (namely
$\textit{sd}^{\: g}$) and best response (namely $\textit{br}^{\: g}$)
in general transfinite iterations (i.e., iterations beyond $\omega_0$)
of the corresponding operator are necessary to reach the outcome. So
to achieve equalities in the above theorem transfinite iterations of
the $T_{\overline{\phi}}$ operator are necessary.

By instantiating $\phi_i$s to specific properties we get
instances of the above result that relate to specific definitions
of rationality. Before we do this we establish another result that
will apply to another class of properties $\phi_i$.

Consider the following natural property
of the underlying functions $s_i(\cdot)$s:

\begin{description}

\item[A] For each strategy $s_i$ from $H$ a state $\omega \in
\Omega$ exists such that $s_i = s_i(\omega)$.

\end{description}

\begin{theorem} \label{thm:epist}
Suppose that property \textbf{A} holds and
\begin{equation}
  \label{equ:1}
\mbox{$\phi_i(s_i, (\C{s_1}, \LL, \C{s_n}))$ for all $i \in [1..n]$ and $s_i \in T_i$.}  
\end{equation}
Then 
\[
G_{CK(\overline{\phi})} = G_{CB(\overline{\phi})} = H.
\]
\end{theorem}
\Proof
As noted in the proof of Theorem \ref{thm:epist1}, for all $\overline{\phi}$
we have $G_{CK(\overline{\phi})} \sse G_{CB(\overline{\phi})}$.
So it suffices to prove that $H \sse G_{CK(\phi)}$.

So take a strategy $s_i$ of player $i$ in $H$. By property \textbf{A} a state
$\omega$ exists such that $s_i = s_i(\omega)$.
Choose for each player $j$ a knowledge correspondence $P_j$
such that $P_j(\omega) = \C{\omega}$.
Then 
\[
G_{P_j(\omega)} = (\C{s_1(\omega)}, \LL, \C{s_n(\omega)})
\]
and, on the account of (\ref{equ:1}),
each player $j$ is $\phi_j$-rational in $\omega$. 

By the choice of the knowledge correspondences
$\C{\omega}$ is an
evident event. Hence it is common knowledge in $\omega$ that each
player $j$ is $\phi_j$-rational in $\omega$.  So by definition $s_i$ is an
element of the $i$th component of $CK_{\phi}$.  
\HB
\VV

Note that any property $\phi_i$ that satisfies (\ref{equ:1})
and is not trivial (that is, for some strategy $s_i$, $\phi_i(s_i, H)$ does not hold)
is not monotonic.

\section{$\LL$ and their consequences}
\label{sec:consequences}

Let us analyze now the consequences of the above two theorems.
Consider first Theorem \ref{thm:epist1}.
The following lemma, in which we refer to the properties introduced
in Section \ref{sec:setup}, clarifies the matters.

\begin{lemma} \label{lem:mono}
The properties 
$\textit{sd}^{\: g}, \ \textit{msd}^{\: g}$ and $\textit{br}^{\: g}$
are monotonic.
\end{lemma}
\Proof
Straightforward.
\HB
\VV

So Theorem \ref{thm:epist1} applies to the above three properties.
(Note that $\textit{br}^{\: g}$ actually comes in three 'flavours'
depending on the choice of beliefs.) Strict dominance in the
sense of $\textit{sd}^{\: g}$ is studied in \cite{CLL05}, while
$\textit{br}^{\: g}$ corresponds to the rationalizability notion of
\cite{Ber84}.

To see the consequences of Theorem \ref{thm:epist}
note the following lemma.
\begin{lemma}
The properties 
$\textit{sd} ^{\: l}, \ \textit{msd} ^{\: l}$ and $\textit{br} ^{\: l}$
satisfy (\ref{equ:1}).
\end{lemma}
\Proof
Straightforward.
\HB
\VV

So Theorem \ref{thm:epist} shows that the `customary' concepts of
strict dominance, $\textit{sd} ^{\: l}$ and $\textit{msd} ^{\: l}$ and
the 'local' version of the best response property $\textit{br} ^{\:
  l}$ cannot be justified in the used epistemic framework as `stand
alone' concepts of rationality.  Indeed, this theorem shows that
common knowledge that each player is rational in one of these three
senses does not exclude any strategy.

What \emph{can} be done is to justify these concepts as
\emph{consequences} of the common knowledge of rationality defined in
terms of $\textit{br} ^{\: g}$, the `global' version of the best
response property, Namely, we have the following result. When each
property $\phi_i$ equals $\phi$, we write here $CK({\phi})$ instead of
$CK({\overline{\phi}})$ and analogously for $CB$.

\begin{theorem} \label{thm:just}
For all games $H$
\[
G_{CK(\textit{br}^{\: g})} = G_{CB(\textit{br}^{\: g})} \sse T^{\infty}_{\textit{sd}^{\: l}},
\]
where we take as the set of beliefs the set of joint strategies of the opponents.
\end{theorem}

\Proof
By Lemma \ref{lem:mono} and Theorem \ref{thm:epist1} 
$G_{CK(\textit{br}^{\: g})} = G_{CB(\textit{br}^{\: g})} = T^{\infty}_{\textit{br}^{\: g}}$.
Each best response to a joint strategy of the opponents is not
strictly dominated, so for all restrictions $G$
\[
T_{\textit{br}^{\: g}}(G) \sse T_{\textit{sd}^{\: g}}(G)
\]
and also 
\[
T_{\textit{sd}^{\: g}}(G) \sse T_{\textit{sd}^{\: l}}(G).
\]
So by Lemma \ref{lem:inc} 
$T^{\infty}_{\textit{br}^{\: g}} \sse
T^{\infty}_{\textit{sd}^{\: l}}$, which concludes the proof.
\HB
\VV

The above result formalizes and justifies in the epistemic
framework used here the often used statement:
\begin{quote}
  common knowledge of rationality implies that the players will choose
  only strategies that survive the iterated elimination of strictly
  dominated strategies
\end{quote}
for games with \emph{arbitrary strategy sets}
and \emph{transfinite iterations} of the
elimination process.  

In the case of finite games we have the following known result.
For a proof using Harsanyi type spaces see \cite{BF06}.

\begin{theorem} \label{thm:just1}
For all finite games $H$

$
G_{CK(\textit{br}^{\: g})} = G_{CB(\textit{br}^{\: g})} \sse T^{\infty}_{\textit{msd}^{\: l}},
$

where we take as the set of beliefs the set of joint mixed strategies
of the opponents.
\end{theorem}

\Proof
The argument is analogous as in the previous proof
but relies on a subsidiary result.

Again by Lemma \ref{lem:mono} and Theorem \ref{thm:epist1} 
$G_{CK(\textit{br}^{\: g})} =  G_{CB(\textit{br}^{\: g})} =
T^{\infty}_{\textit{br}^{\: g}}$.
Further, for all restrictions $G$
\[
T_{\textit{br}^{\: g}}(G) \sse T_{\textit{br}^{\: l}}(G)
\]
and
\[
T_{\textit{br}^{\: l}}(G) \sse T_{\textit{brc}^{\: l}}(G),
\]
where ${brc}^{\: l}$ stands for the best response property
w.r.t.~the correlated strategies of the opponents.  So by Lemma
\ref{lem:inc} $T^{\infty}_{\textit{br}^{\: g}} \sse
T^{\infty}_{\textit{brc}^{\: l}}$.

But by the result of \cite[ page 60]{OR94} (that is a modification of
the original result of \cite{Pea84}) for all restrictions $G$ we have
$T_{\textit{brc}^{\: l}}(G) = T_{\textit{msd}^{\: l}}(G)$, so
$T^{\infty}_{\textit{brc}^{\: l}} = T^{\infty}_{\textit{msd}^{\: l}}$,
which yields the conclusion.  
\HB 
\VV

%The second inclusion justifies this statement and the corresponding
%statement for common belief for finite games and
%strict dominance by mixed strategies, a result well-known in the
%literature, see, e.g. Proposition 3.10 in \cite{BB99}.

\section*{Acknowledgements}

We acknowledge helpful discussions with Adam Brandenburger, who
suggested Theorems \ref{thm:just} and \ref{thm:just1}, and with
Giacomo Bonanno who, together with a referee of \cite{Apt07} suggested
to incorporate common beliefs in the analysis.  Joe Halpern pointed us
to \cite{MS89}.  Jonathan Zvesper provided helpful comments on the
paper.

\bibliography{/ufs/apt/bib/e,/ufs/apt/bib/apt}
\bibliographystyle{handbk}

\end{document}
\section*{Appendix}

\textbf{Proof of Lemma \ref{lem:inc}.}

\NI
We first prove by transfinite induction that for all $\alpha$
\begin{equation}
T_1^{\alpha} \sse T_2^{\alpha}.
  \label{equ:inc}
\end{equation}

By the definition of the iterations we only need to consider the induction
step for a successor ordinal.  So suppose the claim holds for some
$\alpha$. Then by the first two assumptions and the induction
hypothesis we have the following string of inclusions and equalities:
\[
T_1^{\alpha + 1} =   T_1(T_1^{\alpha}) \sse T_1(T_2^{\alpha}) \sse T_2(T_2^{\alpha}) = T_2^{\alpha + 1}.
\]

This shows that for all $\alpha$ (\ref{equ:inc}) holds.
By Tarski's Fixpoint Theorem and Note \ref{note:con} the outcomes of
$T_1$ and $T_2$ exist, which implies the claim.
\HB

\VV

\NI
\textbf{Proof of Theorem \ref{thm:epist1}.}

\NI
We prove three inclusions.

\NI
(i) $G_{CK(\overline{\phi})} \sse G_{CB(\overline{\phi})}$.
\II

This inclusion (for an arbitrary $\overline{\phi}$)
is an immediate consequence of the following alternative
characterization of common knowledge due to \cite{MS89}: if each $P_i$
is a knowledge correspondence, an event $E$ is a common knowledge in
the state $\omega \in \Omega$ if for some evident event $F$ we have
$\omega \in F \sse K E$, where $K E = \{\omega \in \Omega \mid \fa i
\in [1..n] \: P_i(\omega) \sse E\}$.
\II

\newpage

\NI
(ii) $G_{CB(\overline{\phi})} \sse T_{\overline{\phi}}^{\infty}$.
\II

Take a strategy $s_i$ that is an element of the $i$th component of 
$G_{CB(\overline{\phi})}$. So $s_i = s_i(\omega)$ for some
$\omega \in {CB(\overline{\phi})}$.
Then $\omega \in \textbf{RAT}({\overline{\phi}})$ and $\omega \in B^* \textbf{RAT}({\overline{\phi}})$.
The latter implies that for some evident event $F$
\begin{equation}
  \label{equ:F}
\omega \in F \sse \{\omega' \in \Omega \mid \fa i \in [1..n] \: P_i(\omega') \sse \textbf{RAT}({\overline{\phi}})\}.  
\end{equation}

Take now an arbitrary $\omega' \in F \cap \textbf{RAT}({\overline{\phi}})$ and $i \in [1..n]$.
Since $\omega' \in \textbf{RAT}({\overline{\phi}})$, 
player $i$ is $\phi_i$-rational in $\omega'$,
i.e., $\phi_i(s_i(\omega'), G_{P_i(\omega')})$ holds.
But $F$ is evident, so $P_i(\omega') \sse F$. Moreover by (\ref{equ:F})
$P_i(\omega') \sse \textbf{RAT}({\overline{\phi}})$, so 
$P_i(\omega') \sse F \cap \textbf{RAT}({\overline{\phi}})$.
Hence 
$G_{P_i(\omega')} \sse G_{F \cap \textbf{RAT}({\overline{\phi}})}$ and
by the monotonicity of $\phi_i$ we
conclude that $\phi_i(s_i(\omega'), G_{F \cap \textbf{RAT}({\overline{\phi}})})$
holds.

By the definition of $T_{\overline{\phi}}$
this means that $G_{F \cap \textbf{RAT}({\overline{\phi}})} \sse T_{\overline{\phi}}(G_{\textbf{RAT}({\overline{\phi}})})$, i.e. that
$G_{F \cap \textbf{RAT}({\overline{\phi}})}$ is a post-fixpoint of $T_{\overline{\phi}}$.
Hence by Tarski's Fixpoint Theorem 
$G_{F \cap  \textbf{RAT}({\overline{\phi}})} \sse T_{\overline{\phi}}^{\infty}$.

But $s_i = s_i(\omega)$ and $\omega \in F \cap {\textbf{RAT}({\overline{\phi}})}$, so 
we conclude by the above inclusion 
that $s_i$ is an element of the $i$th component of $T_{\overline{\phi}}^{\infty}$.
This proves $G_{CB(\overline{\phi})} \sse T_{\overline{\phi}}^{\infty}$.
\II

\NI
(iii) $T_{\overline{\phi}}^{\infty} \sse G_{CK(\overline{\phi})}$.

Recall that 
$H = (T_1, \LL, T_n, p_1, \LL, p_n)$. We first define 
\begin{itemize}
\item the functions $s_1:\Omega \myra T_1, \LL, s_n:\Omega \myra T_n$,

\item an event $E$,

\item the knowledge correspondences $P_1, \LL, P_n$.

\end{itemize}

Suppose $T^{\infty}_{\overline{\phi}} = (S_1, \LL, S_n)$. Choose $j \in
[1..n]$ such that the set $S_{j_0}$ has the largest cardinality
among the sets $S_1, \LL, S_n$. Define the
function $s_{j_0}:\Omega \myra T_{j_0}$ arbitrarily, but so that it
is onto (note that this is possible since by assumption
$|\Omega| \geq |T_{j_0}|$) and let $E :=
s^{-1}_{j_0}(S_{j_0})$.

Our aim is to ensure that 
\[
G_E = T_{\overline{\phi}}^{\infty}.
\]
So we define each function $s_k: \Omega \myra T_k$, where $k \neq j_0$, in such a way
that $s^{-1}_{k}(S_{k}) = E$. Note that this is possible since 
$|E| \geq |S_{j_0}| \geq |S_k|$.

Next, we define each knowledge correspondence $P_i$
arbitrarily but so that for all $\omega \in E$ we have $P_i(\omega) =
E$.  Then for all $i \in [1..n]$
\[
G_{P_i(\omega)} = G_E.
\]

We now show that for all $\omega \in E$ each player $i$ is
$\phi_i$-rational in $\omega$.  So take an arbitrary $\omega \in E$
and $i \in [1..n]$.  By the definition of the function $s_i(\cdot)$ a
strategy $s_i \in S_i$ exists such that $s_i = s_i(\omega)$.  Now,
$T_{\overline{\phi}}^{\infty}$ is a fixpoint of $T_{\overline{\phi}}$,
so $\phi_i(s_i, T_{\overline{\phi}}^{\infty})$ holds.  But
$T_{\overline{\phi}}^{\infty} = G_E = G_{P_i(\omega)}$, so
$\phi_i(s_i(\omega), G_{P_i(\omega)})$ holds, i.e. player $i$ is
indeed $\phi_i$-rational in $\omega$.

To complete the proof take now an arbitrary strategy $s_i \in S_i$. By
the definition of the function $s_i(\cdot)$ a state $\omega \in E$
exists such that $s_i = s_i(\omega)$.  Further, we just showed that 
each player $j$ is
$\phi_j$-rational in $\omega$.  But by the definition of the
knowledge correspondences
$E$ is an evident event, so it is common
knowledge in $\omega$ that each player $j$ is $\phi_j$-rational in
$\omega$. Hence $\omega \in CK(\overline{\phi})$ and consequently
$s_i$ is an element of the $i$th component of
$G_{CK(\overline{\phi})}$.

This proves that 
$T_{\overline{\phi}}^{\infty} \sse G_{CK(\overline{\phi})}$.
\HB
\VV

\NI
\textbf{Proof of Theorem \ref{thm:epist}.}

\NI
As noted in the proof of Theorem \ref{thm:epist1}, for all $\overline{\phi}$
we have $G_{CK(\overline{\phi})} \sse G_{CB(\overline{\phi})}$.
So it suffices to prove that $H \sse G_{CK(\phi)}$.

So take a strategy $s_i$ of player $i$ in $H$. By property \textbf{A} a state
$\omega$ exists such that $s_i = s_i(\omega)$.
Choose for each player $j$ a knowledge correspondence $P_j$
such that $P_j(\omega) = \C{\omega}$.
Then 
\[
G_{P_j(\omega)} = (\C{s_1(\omega)}, \LL, \C{s_n(\omega)})
\]
and, on the account of (\ref{equ:1}),
each player $j$ is $\phi_j$-rational in $\omega$. 

By the choice of the knowledge correspondences
$\C{\omega}$ is an
evident event. Hence it is common knowledge in $\omega$ that each
player $j$ is $\phi_j$-rational in $\omega$.  So by definition $s_i$ is an
element of the $i$th component of $CK_{\phi}$.  
\HB
\end{document}

%% file: ecmds.tex
\newcommand{\oldbfe}[1]{\begin{bfseries}\emph{#1}\end{bfseries}}

\newcommand{\ES}{\mbox{$\emptyset$}}

\newcommand{\myra}{\mbox{$\:\rightarrow\:$}}

\newcommand{\sse}{\mbox{$\:\subseteq\:$}}

\newcommand{\fa}{\mbox{$\forall$}}
\newcommand{\te}{\mbox{$\exists$}}

\newcommand{\LL}{\mbox{$\ldots$}}

\newcommand{\C}[1]{\mbox{$\{{#1}\}$}}           % curly braces

\newcommand{\NI}{\noindent}
\newcommand{\HB}{\hfill{$\Box$}}
\newcommand{\VV}{\vspace{5 mm}}

\newcommand{\II}{\vspace{2 mm}}

\newcommand{\szkew}[1]{\relax \setbox0=\hbox{\kern -24pt $\displaystyle#1$\kern 0pt }%
\box0}
{\catcode`\@=11 \global\let\ifjusthvtest@=\iffalse}
\newcounter{oldmycaption}